\documentclass[12pt,letter]{article}

\usepackage{graphicx, epsfig, color, cite}
\def\eslt{\not\!\!{E_T}}
\def\esl{\not\!\!{E}}
\def\to{\rightarrow}

\def\bi{\begin{itemize}}
\def\ei{\end{itemize}}
\def\te{\tilde e}

\def\tst{\tilde t}
\def\ttau{\tilde \tau}

\def\tg{\tilde g}
\def\tnu{\tilde\nu}
\def\tell{\tilde\ell}
\def\tq{\tilde q}
\def\tw{\widetilde W}
\def\tz{\widetilde Z}
\def\alt{\stackrel{<}{\sim}}
\def\agt{\stackrel{>}{\sim}}
\def\be{\begin{equation}}  
\def\ee{\end{equation}}  

\newcommand\prd[3]{{\it Phys.\ Rev.\ }{\bf D #1} (#2) #3}
\newcommand\prep[3]{{\it Phys.\ Rept.\ }{\bf #1} (#2) #3}
\newcommand\prl[3]{{\it Phys.\ Rev.\ Lett.\ }{\bf #1} (#2) #3}
\newcommand\plb[3]{{\it Phys.\ Lett.\ }{\bf B #1} (#2) #3}

\newcommand\jhep[3]{{\it J. High Energy Phys.\ }{\bf #1} (#2) #3}

\newcommand\ijmpa[3]{{\it Int.\ J.\ Mod.\ Phys.\ }{\bf A #1} (#2) #3}
\newcommand\npb[3]{{\it Nucl.\ Phys.\ }{\bf B #1} (#2) #3}

\newcommand\zpc[3]{{\it Z.\ Physik }{\bf C #1} (#2) #3}
\newcommand{\hepph}[1]{hep-ph/#1}

\begin{document}
\begin{titlepage}

\begin{flushright}

\end{flushright}

\vspace*{0.5cm}
\begin{center}
{\Large \bf 
Testing the gaugino AMSB model \\
at the Tevatron via slepton pair production
}\\
\vspace{1.2cm} \renewcommand{\thefootnote}{\fnsymbol{footnote}}
{\large Howard Baer$^{1}$\footnote[1]{Email: baer@nhn.ou.edu },
Senarath de Alwis$^2$\footnote[2]{Email: dealwiss@colorado.edu}, 
Kevin Givens$^2$\footnote[3]{Email: kevin.givens@colorado.edu},\\
Shibi Rajagopalan$^1$\footnote[3]{Email: shibi@nhn.ou.edu},
Warintorn Sreethawong$^1$\footnote[4]{Email: wstan@nhn.ou.edu}} \\
\vspace{1.2cm} \renewcommand{\thefootnote}{\arabic{footnote}}
{\it 
1. Dept. of Physics and Astronomy,
University of Oklahoma, Norman, OK 73019, USA \\
2. Dept. of Physics,
University of Colorado, Boulder CO 80309, USA \\
}

\end{center}

\vspace{0.5cm}
\begin{abstract}
\noindent 
Gaugino AMSB models-- wherein scalar and trilinear soft SUSY breaking terms are
suppressed at the GUT  scale while gaugino masses adopt the AMSB form--
yield a characteristic SUSY particle mass spectrum with light sleptons 
along with a nearly degenerate wino-like lightest neutralino 
and quasi-stable chargino.
The left- sleptons and sneutrinos can be pair produced at sufficiently 
high rates to yield observable signals at the Fermilab Tevatron.
We calculate the rate for isolated single and dilepton plus missing energy
signals, along with the presence of one or two highly ionizing chargino tracks.
We find that Tevatron experiments should be able to probe gravitino masses
into the $\sim 55$ TeV range for  inoAMSB models, which corresponds
to a reach in gluino mass of over 1100 GeV.

\vspace{0.8cm}

\noindent PACS numbers: 14.80.Ly, 12.60.Jv, 11.30.Pb

\end{abstract}

%12.60.Jv   Supersymmetric models
%14.80.Ly   Supersymmetric partners
%11.30.Pb   Supersymmetry

\end{titlepage}

\section{Introduction}
\label{sec:intro}

Searches for supersymmetry (SUSY) at the Fermilab Tevatron collider usually
focus on gluino and squark pair production reactions, due to their large
strong interaction production rates\cite{tevgsq,cdf_gsq,d0_gsq}, or on 
observation of chargino-neutralino production and decay to isolated trileptons,
due to their low background rates\cite{trilep,cdf_3l,d0_3l}.
The possibility of observation of slepton pair production at the Tevatron 
was examined in Ref. \cite{slep} in the context of the MSSM with gaugino mass
unification and found to be difficult: 
the dilepton signature from $p\bar{p}\to\tell^+\tell^-\to \ell^+\ell^-+\eslt$ 
(here, $\ell =e$ or $\mu$) is beset with large backgrounds from $W^+W^-$ and $Z\to\tau^+\tau^-$ production, 
while the $\tell\tnu_L\to\ell^\pm +\eslt$ signal is beset by even larger backgrounds
from direct $W^\pm\to\ell^\pm\nu_\ell$ production. However, these past works did not anticipate
the Tevatron reaching integrated luminosities in the vicinity of 8-16 fb$^{-1}$. 

In this paper, we investigate the recently introduced {\it gaugino AMSB}
model (inoAMSB)\cite{inoamsb}, which arises naturally from some highly motivated 
string theory constructions. The inoAMSB model gives rise to a characteristic
SUSY particle mass spectrum which features 1. a wino-like lightest neutralino
$\tz_1$, 2. a nearly mass degenerate quasi-stable chargino $\tw_1$,
(points 1 and 2 also occur in previous AMSB constructs\cite{amsb,ggw,amsb_col}), 
3. a rather light spectrum of
sleptons, arranged in a mass hierarchy $m_{\tnu_L}<m_{\tell_L}<m_{\tell_R}$ and 
4. a rather heavy spectrum of squarks and gluinos, where 
$m_{\tg}\sim m_{\tq}\sim 7.5 m_{\tw_1}$. 
Given the LEP2 limit on quasi-stable
charginos from AMSB models, where $m_{\tw_1}>91.9$ GeV\cite{lep2}, this implies
$m_{\tg}\agt 700$ GeV: quite beyond the reach of Tevatron. 
However, in inoAMSB models
the sleptons can exist with masses as low as $\sim 130$ GeV. 
Pair production of inoAMSB sleptons, followed by decays into quasi-stable charginos, 
should give rise to characteristic isolated single or dilepton plus $\eslt$
signatures, accompanied by the presence of one or two highly ionizing chargino
tracks (HITs)\cite{ggw}. 

In a previous work\cite{inoamsb}, we presented the spectrum of SUSY particle masses
which are expected from inoAMSB models, and evaluated prospects for 
detection at the LHC with $\sqrt{s}=14$ TeV. A 100 fb$^{-1}$ LHC reach
to $m_{\tg}\sim 2.3$ TeV was found. The gluino and squark cascade decay\cite{cascade}
events would often contain the presence of highly ionizing chargino tracks
that could range up to a few {\it cm} in length. The unique inoAMSB mass spectrum
$m_{\tz_2}>m_{\tell_R}>m_{\tell_L}>m_{\tw_1 , \tz_1}$ leads
to a characteristic {\it double bump} (mass edge) structure in the
opposite-sign dilepton invariant mass distribution which could serve to distinguish
the inoAMSB model from minimal AMSB (mAMSB) or hypercharged AMSB\cite{hcamsb} (HCAMSB). 

In Ref. \cite{dm},the relic density of dark matter in inoAMSB (and also in mAMSB and 
HCAMSB) was considered. In all AMSB models with sub-TeV scale $\tz_1$, the
thermal abundance of neutralino cold dark matter is well below the 
WMAP-measured value of $\Omega_{CDM}h^2=0.1123\pm 0.0035$\cite{wmap7}.
However, the possibility of additional neutralino production via moduli\cite{mr}, 
gravitino\cite{gravdec} or axino\cite{axdec} decay can augment the thermal abundance, bringing the
expected neutralino abundance into accord with measured values.

In this paper, we calculate signal rates for slepton pair production
in inoAMSB models at the Fermilab Tevatron collider. We find a considerable reach 
for the nearly background free signature of single or OS dilepton plus $\eslt$
plus one or two HITs\cite{ggw}; these signal rates ought to allow Tevatron experiments
to explore slepton masses from the inoAMSB model into the 200 GeV range
for $\sim 10$ fb$^{-1}$ of integrated luminosity, corresponding to a reach
in $m_{3/2}$ of over 50 TeV.

\section{The gaugino AMSB model}
\label{sec:inoAMSB}

Gaugino Anomaly Mediated Supersymmetry Breaking\cite{inoamsb} is a very simple 
scenario for generating  SUSY breaking soft terms in 
low energy supersymmetric theories. 
The main assumption is that the high energy theory which generates SUSY breaking 
is of the sequestered type \cite{amsb}, which effectively means that the classical gaugino and 
scalar masses and $A$-terms are highly suppressed relative to the gravitino mass scale. 
This is in contrast to the situation in usual supergravity (SUGRA) models, 
where these soft parameters are classically generated at the gravitino mass scale. 
Nevertheless, in contrast to what is usually advocated in AMSB \cite{amsb}, 
it has been argued \cite{Kaplunovsky:1993rd} that only {\it gaugino masses} 
are generated by Weyl anomalies. 
In inoAMSB\cite{inoamsb}\cite{deAlwis:2009tp}, the scalar masses are then 
generated by renormalization group (RG) running as in what is often called 
gaugino mediation\cite{inomsb} or simple no-scale SUSY breaking models\cite{noscale}. 
The inoAMSB model then avoids both the generic FCNC problems of gravity mediated 
scenarios and also the tachyonic slepton problem of the traditional AMSB construct.
It also avoids the presence of tau slepton LSPs which occur in 
gaugino mediation/no-scale models with gaugino masses unified at a high scale.

This very simple phenomenological model depends on just two parameters: 
the gravitino mass $m_{3/2}$ which sets the scale for all sparticle masses, 
and $\tan\beta$, the ratio of the the Higgs vacuum expectation values in the MSSM.  
In fact, it appears to be the simplest SUSY mediation model that one can conceive of 
which satisfies all phenomenological constraints. 

Furthermore, inoAMSB can be realized within a highly motivated class of 
string theories\cite{deAlwis:2009tp}. 
The models in question are called the large volume compactification scenario (LVS) 
of type IIB string theory and were introduced in \cite{Balasubramanian:2005zx}. 
The moduli (and the dilaton) of string theory, which appear as 4D fields in the 
effective action, are stabilized using a combination of fluxes and 
non-perturbative effects (for reviews see \cite{Grana:2005jc}). 
The Calabi-Yau (CY) manifolds on which the theory is compactified to 4D is of the 
so-called ``Swiss Cheese" type with one large four cycle 
(which controls the overall size of the internal space) and one or more small cycle. 
An analysis of the potential for the moduli shows that the volume is exponentially 
large in the small cycle(s) whose size in turn is stabilized at values 
larger than the string scale. 
The effective parameter which controls this is determined by the 
Euler character of the CY manifold and the (flux dependent) value of the dilaton.

It was shown in \cite{deAlwis:2009tp} that in these models, 
for large enough volume (greater than $10^5$ Planck units), FCNC effects are suppressed. 
Indeed, all classically generated soft SUSY breaking parameters  are volume 
suppressed compared to the gaugino mass soft terms that are generated by anomaly 
mediation. The latter effect  is actually a consequence of the generation of 
gaugino masses by the Weyl anomaly effect as discussed in \cite{Kaplunovsky:1993rd}. 

The phenomenology of this class of string theoretic models is  effectively 
controlled by the gravitino mass. But  the theory at this point only allows 
us to estimate an upper bound to the possible size of $\mu$ and $B$ terms. 
So we use the latter after trading it  for (as is usual)  $\tan\beta$, 
and regard the former as an output from the experimental value of the $Z$ mass. 
The parameters of the phenomenological model which comes from these 
string theory considerations are thus
\be
m_{3/2},\ \tan\beta ,\ sign(\mu ) .
\label{eq:pspace}
\ee

The gravitino mass determines the values of the gaugino masses at the high scale 
(which will be chosen to be the GUT scale) by the Weyl anomaly formula 
given in \cite{Kaplunovsky:1993rd}. 
It turns out that for the scenario in \cite{deAlwis:2009tp}, 
this is exactly the same as what is often given as the AMSB formula for these masses 
{\it i.e.}
\be
M_i= \frac{b_i g_i^2}{16\pi^2}m_{3/2},
\ee
with $b_i=(33/5,\ 1,\ -3)$.
The initial (high scale) values of the other soft parameters are then taken to be 
\be
m_0=A_0=0, 
\ee
where $m_0$ is the common soft SUSY breaking scalar mass evaluated at 
the high scale $\sim M_{string}$ or $M_{GUT}$, 
and $A_0$ is the trilinear soft SUSY breaking (SSB) term.

\section{Production and decay of inoAMSB sleptons at the Tevatron}
\label{sec:prod}

We begin by examining the sort of sparticle mass spectra that is
expected from  the inoAMSB boundary conditions: $m_0=A_0=0$ but
with $M_i=\frac{b_i g_i^2}{16\pi^2}m_{3/2}$.
We adopt a unified value of the gauge coupling $g_{GUT}=0.714$ and then 
for a given value of $m_{3/2}$ compute the GUT scale
values of the three gaugino masses $M_i$ for $i=1-3$.
We compute the sparticle mass spectra using the Isasugra subprogram of the
event generator Isajet\cite{isajet}, along with the option of
non-universal gaugino masses. 
The parameter space is that of Eq. \ref{eq:pspace}.

After input of the above parameter set, Isasugra
implements an iterative procedure of solving the MSSM RGEs for the
26 coupled renormalization group equations, taking the weak scale
measured gauge couplings and third generation Yukawa couplings as inputs, 
as well as the above-listed GUT scale SSB terms. 
Isasugra implements full 2-loop RG running in the $\overline{DR}$ scheme, 
and minimizes the RG-improved 1-loop effective
potential at an optimized scale choice $Q=\sqrt{m_{\tst_L}m_{\tst_R}}$
(which accounts for leading two-loop terms)\cite{hh}
to determine the magnitude of $\mu$ and the value of $m_A$. 
All physical sparticle masses are computed with complete 1-loop corrections, 
and 1-loop weak scale threshold corrections
are implemented for the $t$, $b$ and $\tau$ Yukawa couplings\cite{pbmz}. 
The off-set of the weak scale boundary conditions due to threshold corrections 
(which depend on the entire superparticle mass spectrum), 
necessitates an iterative up-down RG running solution.
The resulting superparticle mass spectrum is typically in close accord 
with other sparticle spectrum generators\cite{kraml}.

\begin{figure}[t]
\begin{center}
\epsfig{file=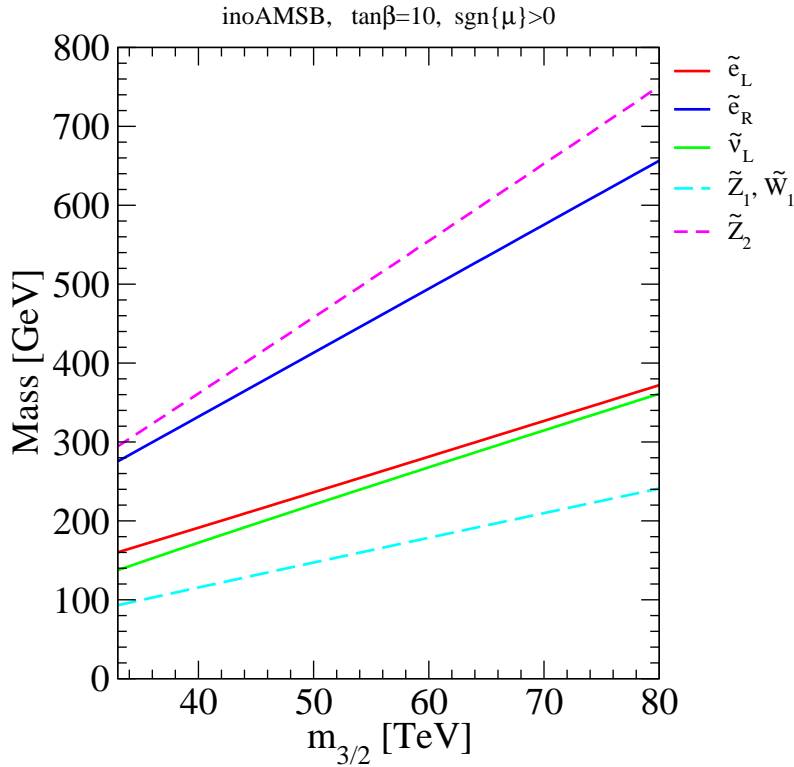,width=12cm,angle=-90}
\end{center}
\vspace*{-4mm}
\caption{\it 
Plot of various gaugino and slepton masses in the inoAMSB model
versus $m_{3/2}$ for $\tan\beta =10$ and $\mu >0$.
}\label{fig:mass} 
\end{figure}

In Fig. \ref{fig:mass}, we show the mass spectrum of various sleptons
and light gauginos of interest to Tevatron experiments versus $m_{3/2}$
for $\tan\beta =10$ and $\mu >0$. Results hardly change if we flip the sign of
$\mu$. If $\tan\beta$ is increased, then third generation squark and slepton 
and heavy Higgs masses decrease, while first/second generation slepton masses 
of interest here remain nearly the same. We see from Fig. \ref{fig:mass} that
while charginos and neutralinos are predicted to be the lightest MSSM particles,
$\tell_L$ and $\tnu_L$ are also quite light-- as low as $\sim 130$ GeV-- with
$m_{\tnu_L}<m_{\tell_L}$. Unlike mSUGRA or mAMSB, the $\tell_R$ mass 
is split from $\tell_L$ and quite a bit heavier: at least $280$ GeV in inoAMSB.
The $\tz_2$ is bino-like, with $m_{\tz_2}>m_{\tell_R}$.

\begin{figure}[t]
\begin{center}
\epsfig{file=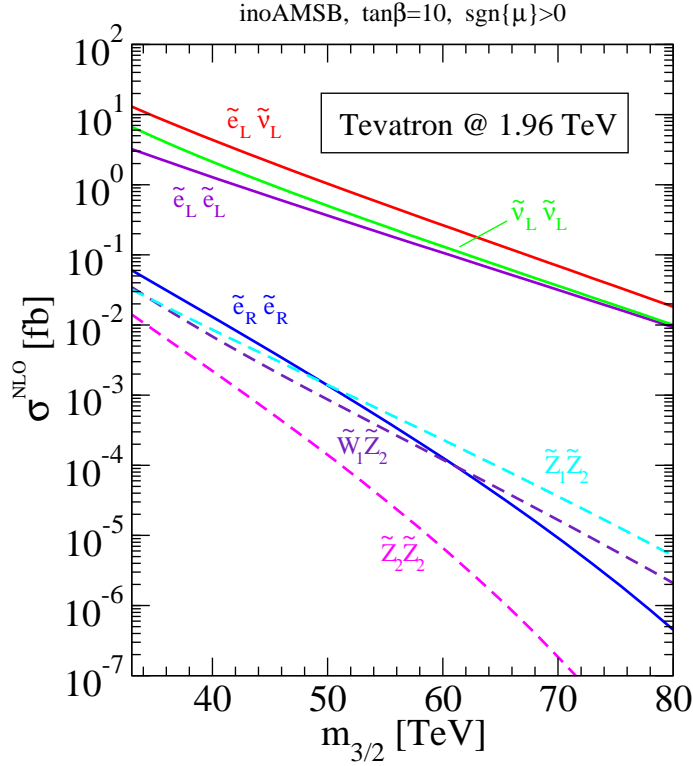,width=12cm,angle=-90}
\end{center}
\vspace*{-4mm}
\caption{\it 
Plot of various slepton and gaugino pair production cross sections
at the Fermilab Tevatron collider with $\sqrt{s}=1.96$ TeV for the inoAMSB model.
We plot versus $m_{3/2}$ for $\tan\beta =10$ and $\mu >0$.
}\label{fig:xs} 
\end{figure}

In Fig. \ref{fig:xs}, we show various slepton pair production cross sections
as calculated at NLO\cite{bhr} using the Prospino program\cite{slnlo,prospino}.\footnote{Recent works on slepton pair production at hadron colliders including
resummation effects are included in Ref. \cite{resum}.}
The results are calculated versus $m_{3/2}$ for the same parameters as in 
Fig. \ref{fig:mass}.
We do not present $\tw_1\tw_1$ or $\tw_1\tz_1$ cross sections, since the
visible energy from quasi-stable $\tw_1\to\pi\tz_1$ decay is insufficient to
trigger on. 

From Fig. \ref{fig:xs}, we see that the reactions $p\bar{p}\to \te_L^{\pm}\tnu_{eL}$,
$p\bar{p}\to\tnu_{eL}\bar{\tnu}_{eL}$ and 
$p\bar{p}\to\te_{L}\bar{\te}_{L}$ are comparable and can exceed the 1 fb level for
$m_{3/2}\alt 45$ GeV.
They reach a maximum value of $\sim 10$ fb for $m_{3/2}\sim 33$ TeV. When we sum over
$\ell =e,\ \mu$ and $\tau$, then the total slepton pair production is even larger.
The $\te_R\bar{\te}_R$ pair production is much lower in rate, and unobservable at
projected Tevatron luminosities. Also, we see that cross sections involving
$\tz_2$ production are much smaller, and won't contribute to the observable
rates. For $m_{3/2}\agt 60$ TeV, the slepton pair production cross sections drop below
the 0.1 fb level, and are likely unobservable at Tevatron.

To determine the slepton pair producton signatures, we must next
calculate their branching fractions\cite{bbkmt}. 
Using Isajet, we find the following values:
\bi
\item $\tnu_{\ell L}\to\tz_1\nu_{\ell}$\ \ \ 33\%,
\item $\tnu_{\ell L}\to\tw_1\ell$\ \ \ 67\%,
\ei
while
\bi
\item $\tell_{L}\to\tz_1 \ell$\ \ \ 33\%,
\item $\tell_{L}\to\tw_1\nu_{\ell}$\ \ \ 67\% .
\ei

Each quasi-stable chargino
gives rise to a HIT, which may be visible in the microvertex tracker.
By combining branching fractions with slepton pair production,
we find the following event topologies. 
\begin{enumerate}
\item $\tell_L\tnu_{\ell L}\to \ell +2\ HITs+\eslt$\ \ \ 45\%,
\item $\tell_L\tnu_{\ell L}\to \ell^+\ell^- +\ HIT +\eslt$,\ \ \ 22\%,
\item $\tell_L\tnu_{\ell L}\to \ell +\eslt$,\ \ \ 10\%,
\item $\tnu_{\ell L}\bar{\tnu}_{\ell L}\to \ell^+\ell^- +2\ HITs+\eslt$,\ \ \ 45\%,
\item $\tnu_{\ell L}\bar{\tnu}_{\ell L}\to \ell +\ HIT+\eslt$,\ \ \ 44\%,
\item $\tell_{L}\bar{\tell}_{L}\to \ell +\ HIT +\eslt$,\ \ \ 44\%,
\item $\tell_{L}\bar{\tell}_{L}\to \ell^+\ell^-  +\eslt$,\ \ \ 10\% .
\end{enumerate}
The $\ell^+\ell^- +\eslt$ topology from reaction 7 will likely be
swamped by $WW$ and  $Z\to\tau^+\tau^-$ backgrounds, while the $\ell^\pm +\eslt$ 
topology from reaction 3 will be buried under $W\to \ell\nu_\ell$ background.
However, the event topologies including HITs should stand out from 
SM background, and furthermore, should signal the presence of the quasi-stable chargino.
We note here that topologies 1 and 2 are unique to $\tell_L\tnu_L$ production, 
while topology 4 is unique to $\tnu_{\ell L}\bar{\tnu}_{\ell L}$ production. 
If a two HIT topology has one of the HITs missed for some reason, it will look
like a single HIT event. But the 2 HIT topologies 1 and 4 are unique in that they
each contain two quasi-stable chargino tracks.
Thus, these
topologies will pinpoint the particular superparticle production mechanism.
Topologies 5 and 6 arise from both $\tnu_L\bar{\tnu}_L$ and $\tell_L\bar{\tell}_L$
production. 

The first/second generation slepton masses and branching fractions listed above are 
largely immune to variations in $\tan\beta$, so 
even if $\tan\beta$ changes over the range $\sim 5-40$ (parameter space maxes out at 
$\tan\beta\sim 42$; see Fig. 5 of Ref.~\cite{inoamsb}), the expected signatures 
are expected to be nearly $\tan\beta$ invariant. As $\tan\beta$ increases, the $\ttau_1$
and $\tnu_{\tau L}$ masses decrease, leading to a somewhat increased rate for production of 
one of two tau leptons plus HITs plus $\eslt$ relative to production of
one or two isolated $\ell$s plus HITs plus $\eslt$.

\section{Signal and background after cuts}
\label{sec:signal}

Once the superparticle mass spectrum and decay branching fractions have been
calculated using Isasugra, the output is fed into Herwig\cite{herwig} for 
event generation using $p\bar{p}$ collisions at $\sqrt{s}=1.96$ TeV.
We adopt the AcerDet toy detector simulation program as well\cite{acerdet}.
We then generate all superparticle production events. A large component from
$\tw_1\tz_1$ and $\tw_1^+\tw_1^-$ production will not provide enough visible energy
for triggers, so we focus instead on slepton pair production, where the signal is 
an opposite-sign/same flavor (OSSF) dilepton pair ($e^+e^-$ or $\mu^+\mu^-$) plus
missing $E_T$ (MET).

\begin{figure}[t]
\begin{center}
\epsfig{file=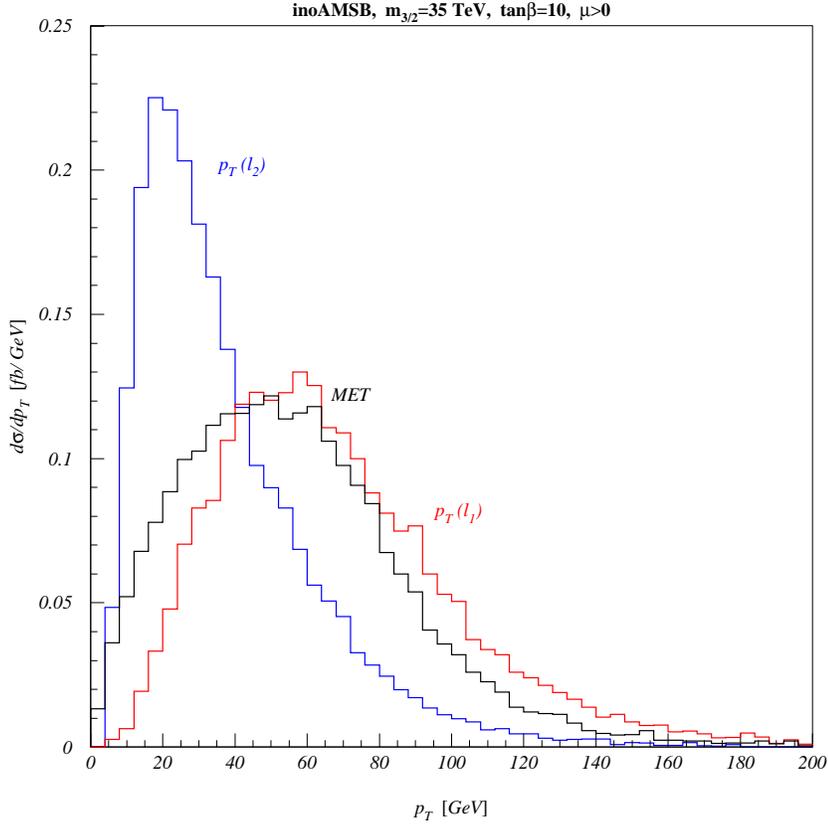,width=12cm,angle=0}
\end{center}
\vspace*{-4mm}
\caption{\it 
Plot of $p_T$ distribution of hardest lepton, second hardest lepton and
$MET$ from slepton pair production events with OSSF dileptons
at the Fermilab Tevatron for the inoAMSB model.
We adopt $m_{3/2}=35$ TeV, $\tan\beta =10$ and $\mu >0$.
}\label{fig:pt} 
\end{figure}

To gain perspective on the energy scales from slepton pair production, we plot first
in Fig. \ref{fig:pt} the $p_T$ distribution of the hardest ($\ell_1$) and softest
($\ell_2$) leptons from slepton pair production in inoAMSB with $m_{3/2}=35$ TeV, 
$\tan\beta =10$ and $\mu >0$. So far, we have imposed no cuts, so the events come from
pure slepton pair production with either one or two isolated leptons in the final state.
The $p_T(\ell_1 )$ distribution spans an 
approximate range $\sim 30-120$ GeV, with a peak at $\sim 65$ GeV. The second lepton
$p_T$ distribution spans $\sim 10-80$ GeV, with a peak at $\sim 20$ GeV. 
We also show the expected MET distribution, which peaks around 60 GeV.

\begin{figure}[t]
\begin{center}
\epsfig{file=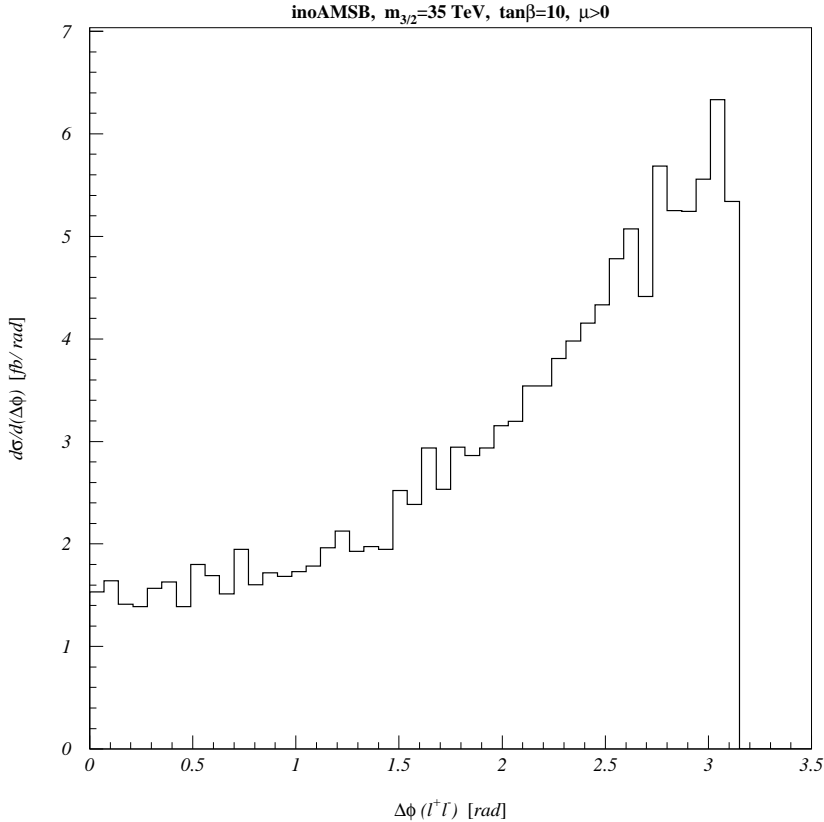,width=12cm,angle=0}
\end{center}
\vspace*{-4mm}
\caption{\it 
Distribution in OSSF dilepton transverse opening angle
at the Fermilab Tevatron for the inoAMSB model.
We adopt $m_{3/2}=35$ TeV, $\tan\beta =10$ and $\mu >0$.
}\label{fig:phi} 
\end{figure}

In Fig. \ref{fig:phi}, we show the OSSF dilepton opening angle in the transverse plane.
The distribution peaks around $\Delta\phi (\ell^+\ell^-)\sim\pi$, reflecting the
fact that the sleptons are produced back-to-back in the transverse direction.
However, when the lepton momentum from slepton decay is boosted to the LAB frame,
the distribution smears out considerably: while most events occur at large
transverse opening angle, there is a significant probability for both
detected leptons to appear on the same side of the detector, {\it i.e.} with
$\Delta\phi (\ell^+\ell^- )<\pi /2$.

Following recent CDF/D0 analyses of $W$ and $Z$ production\cite{d0,cdf}, 
we next impose a minimal set of cuts:
\begin{itemize}
\item $\eslt > 25$ GeV,
\item at least one isolated lepton ($e$ or $\mu$) 
with $p_T(\ell )>25$ GeV and $|\eta (\ell )| <1$,
\item for two lepton events, $p_T(\ell_2 ) >25$ GeV and $|\eta (\ell_2 )| < 2$,
\item for events containing HITs, we require $|\eta (HIT)| < 2$.
\end{itemize}

\begin{figure}[t]
\begin{center}
\epsfig{file=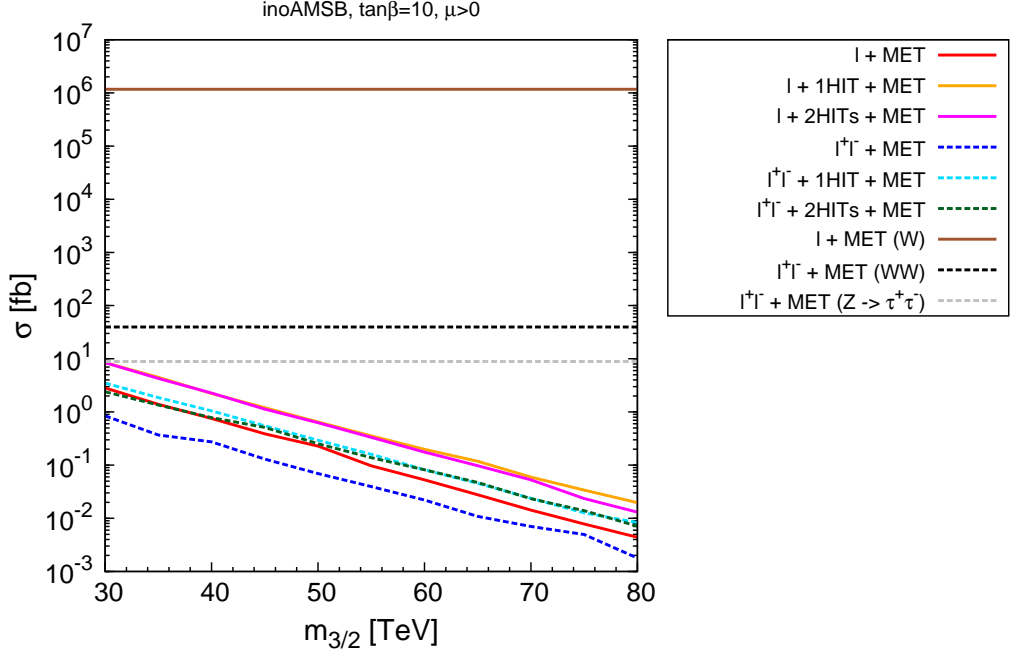,width=14cm,angle=0}
\end{center}
\vspace*{-4mm}
\caption{\it 
Plot of various isolated lepton plus $\eslt$ event topologies after cuts 
at the Fermilab Tevatron for the inoAMSB model.
Here, $\ell =e$ or $\mu$.
We plot signal rate after cuts 
versus $m_{3/2}$ for $\tan\beta =10$ and $\mu >0$.
}\label{fig:signal} 
\end{figure}

Next, keeping $\tan\beta =10$ and $\mu >0$, we scan over $m_{3/2}$ values from
30-80 TeV. The rates for various single and OSSF dilepton events, with 0,1, or 2 HITs,
are shown in Fig. \ref{fig:signal}. We also compute single and OSSF dilepton 
background rates from $p\bar{p}\to W^\pm\to \ell\nu_\ell$ production, and
$W^+W^-$ and $Z\to\tau^+\tau^-$ production, respectively.
The single lepton background from $W$ production is about six orders of magnitude
above signal, making a search in this channel hopeless.
The $WW$ and $\tau^+\tau^-$ backgrounds are somewhat above the largest OSSF dilepton signal levels.

At this stage, it is important to note that most signal events will contain at least one HIT,
which should be well separated in angle from the isolated leptons. The presence of HITs should 
allow distinguishability of signal from background.
The efficiency for HIT identification is detector dependent, and beyond the scope of our 
theory analysis: here we will assume a HIT identification efficiency of 100\%. 
Long-lived tracks from hyperon production with $\Xi\to\Lambda\pi$ decay 
have been identified by the CDF collaboration in the SVX detector 
and used to great effect in their analysis of $\Xi_b$ production and 
decay\cite{cdf_hyp_b}.
If we require the presence of one or more HITs from quasi-stable charginos, then
SM background should be largely negligible. In particular, 
the $\ell^+\ell^- +2\ HITs+\eslt$ signal from sneutrino pair production followed
by $\tnu_\ell\to \ell\tw_1$ decay should provide a smoking gun signature for 
inoAMSB at the Tevatron. From Fig. \ref{fig:signal}, we see that this cross section
ranges up to 2 fb after cuts. With $\sim 10$ fb$^{-1}$ of integrated luminosity,
Tevatron experiments may have a reach for the inoAMSB model in this channel to 
$m_{3/2}\sim 40-50$ TeV. The $1\ell +2\ HITs +\eslt$ channel, coming from
$\te_L\tnu_{\ell L}$ production, 
is generically about a factor 3 higher than the $\ell^+\ell^- +2\ HITs+\eslt$
channel, and should provide corroborating evidence. There are also comparable contributions
to the $\ell +HIT+\eslt$ and $\ell^+\ell^- + HIT +\eslt$ channels. By combining all channels,
the 10 fb$^{-1}$ reach of Tevatron for slepton pair production in inoAMSB models 
should extend to $m_{3/2}\sim 55$ GeV. Augmenting the signal with single tau-jet and ditau-jets
plus $HITs+\eslt$ events will increase the reach  even further. 

Once an inoAMSB signal for slepton pair production is established, then the next step will be to try
to extract sparticle masses from the event kinematics. We will first look at the sneutrino pair
production reaction $p\bar{p}\to\tnu_{\ell L}\bar{\tnu}_{\ell L}\to \ell^+\ell^- +2\ HITs+\eslt$, which arises
when $\tnu_{\ell L}\to \ell\tw_1$ decay. Since the $\tw_1$ gives essentially all $\eslt$-- aside from the HIT-- 
it would be useful to construct the transverse mass\cite{tm} from the $\tnu_{\ell L}$ decay:
\be
m_T^2(\ell ,\vec{\esl}_T)=(|\vec{p}_{\ell T}|+|\vec{\esl}_T |)^2-(\vec{p}_{\ell T}+\vec{\esl}_T)^2
\ee
from each signal event, since this quantity is bounded by 
$m_T(max)=m_{\tnu_{\ell L}}\left(1-m_{\tw_1}^2/m_{\tnu_{\ell L}}^2\right)$. 
However, since we do not {\it a priori} know the
value of $p_T(\tw_1 )$, but only know $\vec{\esl}_T\simeq \vec{p}_T(\tw_1 )+\vec{p}_T(\tw_1')$, we must instead
use the Cambridge $m_{T2}$ variable\cite{mt2}:
\be
m_{T2}={min\atop{\vec{p}_T(\tw_1 )=\vec{\esl}_T-\vec{p}_T(\tw_1')}}\left[max\left(m_T(\ell_1,\vec{p}_T(\tw_1 )),
m_T(\ell_2,\vec{p}_T(\tw_1'))\right)\right]
\ee
which by construction must be bounded by the $m_T$ value which is constructed with the correct lepton and 
missing $E_T$ vectors.

\begin{figure}[t]
\begin{center}
\epsfig{file=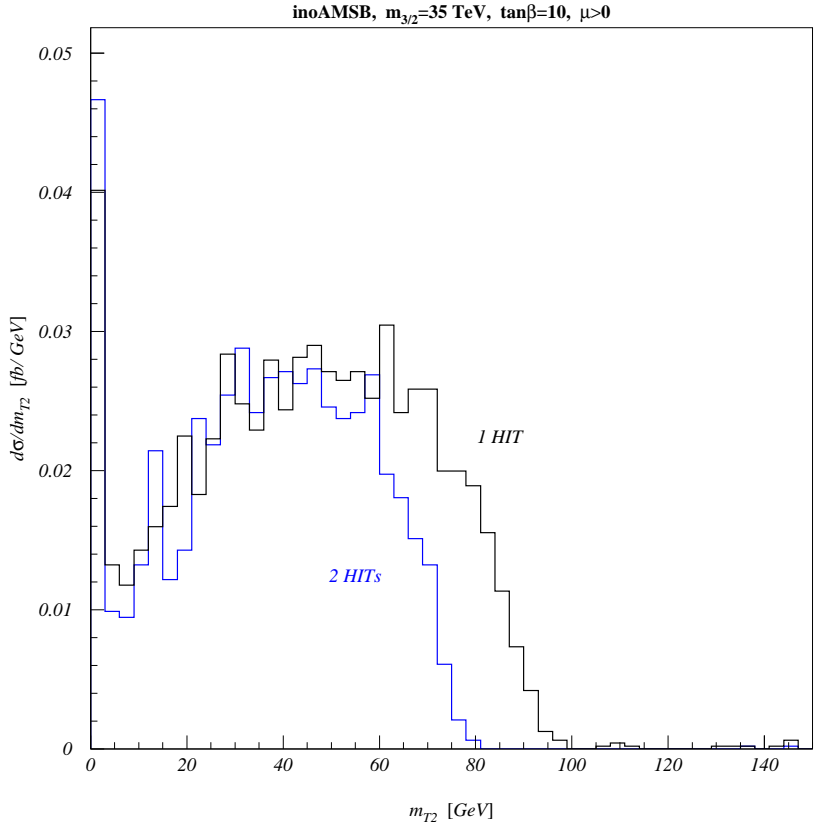,width=12cm,angle=0}
\end{center}
\vspace*{-4mm}
\caption{\it 
Distribution in variable $m_{T2}(\ell^+,\ell^-,\eslt )$
from OSSF slepton pair events at the Tevatron for events
containing 1 or 2 HITs.
We plot for $m_{3/2}=35$ TeV, $\tan\beta =10$ and $\mu >0$.
}\label{fig:mt2} 
\end{figure}

The distribution in $m_{T2}$ for $\ell^+\ell^- +2\ HITs+\eslt$ is shown as the blue histogram in Fig. \ref{fig:mt2}.
We see as expected a continuum distribution followed by a visible cut-off around 
$m_T(max)\simeq 73.4$ GeV.

If instead we examine the $m_{T2}$ distribution for $\ell^+\ell^- +1\ HIT+\eslt$, then we
will mainly pick up $\tell_L^+\tell_L^-$ production, plus some fraction of $\tnu_{\ell L}\bar{\tnu}_{\ell L}$
events where one of the HITs is missed, perhaps due to having too high $|\eta | >2$ value.
In this case, $m_{T2}$ is bounded by 105.9 GeV, as is illustrated in Fig. \ref{fig:mt2}.

\subsection{Slepton pair production in mAMSB}
\label{ssec:mamsb}

We note here that Tevatron experiments can be sensitive to slepton pair production in
the mAMSB model as well\cite{ggw}. Light sleptons occur in mAMSB for very low values of the $m_0$ parameter.
We have examined a case in the mAMSB model with $m_0=220$ GeV, $m_{3/2}=35$ TeV, $\tan\beta =10$ and
$\mu >0$. This mAMSB benchmark gives rise to a spectrum with $m_{\ttau_1}=124$ GeV, $m_{\tnu_{\ell L}}=150$ GeV, 
$m_{\tell_R}=160.3$ GeV,  $m_{\tell_L}=174$ GeV and $m_{\tw_1,\tz_1}\simeq 99.3$ GeV. 
The event rates and distributions are rather similar to the inoAMSB model with $m_{3/2}=35$ TeV.
Naively, one might expect $\tau^+\tau^- +HITs+\eslt$ production to occur at higher rates in
mAMSB than in inoAMSB, since in mAMSB, the $\ttau_1$ is NLSP, while in inoAMSB, the $\tnu_{\ell L}$ is NLSP.
However, since $m_{\tell_R}$ is quite a bit lighter in mAMSB than in inoAMSB, production of 
$\ell^+\ell^- + HITs+\eslt$ is augmented by $\tell_R^+\tell_R^-$ production. Detailed simulations
find a ratio 
\be 
R=\frac{N(\tau^+\tau^- +2\ HITs +\eslt )}{N(\ell^+\ell^- +2\ HITs+\eslt )}
\ee
to be 0.16 for inoAMSB
while $R=0.18$ for mAMSB (here, we require $p_T(\tau -jet )>20$ GeV and $|\eta (\tau -jet)|<2$). 
Thus, it looks difficult to distinguish the two models at the Tevatron
based on slepton pair production. Distinguishing the two models is straightforward once enough
integrated luminosity is accumulated at LHC, since then $\tz_2$s that are produced in gluino and squark 
cascade decays lead to a double edge  structure in the $m(\ell^+\ell^-)$ distribution (reflecting the
large $m_{\tell_L}$, $m_{\tell_R}$ mass gap) while
the mAMSB model with light sleptons gives only a single mass edge, owing to the near degeneracy of
$\tell_R$ and $\tell_L$\cite{inoamsb}.
We also emphasize here that slepton pair production only occurs in mAMSB for very low 
$m_0$ and $m_{3/2}$ values, and the $m_{\tell_{L,R}}-m_{\tz_1}$ mass gap is quite variable for
different $m_0$ values, while in inoAMSB, this mass gap is essentially a fixed prediction
depending only on $m_{3/2}$.

\section{Conclusions}
\label{sec:conclude}

In this paper, we have examined the possibility of detecting slepton pair production from the
gaugino AMSB model at the Fermilab Tevatron, with 10-16 fb$^{-1}$ of integrated luminosity.
This model is characterized by a spectrum of very light sleptons, along with a wino-like
neutralino and a nearly mass degenerate, quasi-stable chargino; the latter occur in most AMSB-type models.
In inoAMSB, the sneutrinos are the lightest sleptons, but they can decay visibly into modes such 
as $\tnu_{\ell L}\to \ell\tw_1$. If the highly ionizing chargino tracks (HITs) can be identified, 
then the $\ell +HITs+\eslt$ and $\ell^+\ell^- +HITs+\eslt$ signatures should be nearly background free.
Summing over all production reactions and final states containing HITs should give the
Fermilab Tevatron a reach in $m_{3/2}$ to $\sim 55$ TeV, which corresponds to a gluino mass
of $\sim 1200$ GeV. This should be somewhat beyond what LHC can explore with $\sqrt{s}=7$ TeV and
$\sim 1$fb$^{-1}$ of integrated luminosity\cite{lhc7}. If a sizable signal is established, then the distribution in
$m_{T2}$ should provide some information on the masses of the sparticles being produced. In particular,
the max of the $m_{T2}$ distribution should be somewhat higher for dilepton events with one HIT, as
opposed to dilepton events containing two HITs. This reflects the $m_{\tell_L}>m_{\tnu_{\ell L}}$
mass hierarchy which is expected from inoAMSB models.

\section*{Acknowledgments}
We thank Yuri Gershtein for discussions.
This work was supported in part by the U.S.~Department of Energy. 
SdA and KG are supported in part by the United States Department of Energy 
under grant DE-FG02-91-ER-40672.

%%%%%%%%%%%%%%%%%%%%%%%%%%%%%%%%%%%%%%%%%%%%%%%%%%%%%%

\end{document}